\documentclass[12pt]{iopart}
\usepackage[dvips]{graphicx}
\usepackage{epsfig}
\usepackage{subfigure}
\usepackage{epsfig}
\usepackage{color}
\newcommand{\bols}{\mathbf{s}}

\newcommand{\btheta}{\mbox{\protect\boldmath $\theta$}}

\newcommand{\order}{{\cal O}}

\newcommand{\hatm}{\hat{m}}
\newcommand{\hatth}{\hat{\theta}}

\newcommand{\tp}{{t^{\prime}}}

\begin{document}
\title[Dynamical TAP equations for non-equilibrium Ising spin glasses]{Dynamical TAP 
equations for non-equilibrium Ising spin glasses}

\author{Yasser Roudi}

\address{Nordita, 106 91 Stockholm, Sweden}
\address{Kavli Institute for Systems Neuroscience, NTNU, 7010 Trondheim, Norway }
\ead{yasser@nordita.org}

\author{John Hertz}
\address{Nordita, 106 91 Stockholm, Sweden}
\address{The Niels Bohr Institute, 2100 Copenhagen, Denmark}
\ead{hertz@nordita.org}
\begin{abstract}
We derive and study dynamical TAP equations for Ising spin glasses obeying both synchronous
and asynchronous dynamics using a generating functional approach. The system can have an asymmetric 
coupling matrix, and the external fields can be time-dependent.  In the synchronously updated model, 
the TAP equations take the form of self consistent equations for magnetizations at time $t+1$, given 
the magnetizations at time $t$. In the asynchronously updated model, the TAP equations determine the 
time derivatives of the magnetizations at each time,
again via self consistent equations, given the current values of the magnetizations. Numerical 
simulations suggest that the TAP equations become exact for large systems.
\end{abstract}

\maketitle
\section{Introduction}
Within the mean field approximation, spin models with quenched disorder can be studied by 
analyzing their quenched averaged behavior or, alternatively, for a specific realization
of the quenched disorder \cite{MezardParisiVirasoro}. In the equilibrium case, the former type of 
analysis includes the replica method, while the latter one is usually formulated as naive mean 
field, TAP equations \cite{Thouless77}, or, more generally, a Plefka expansion \cite{Plefka82}. 
These equations can be derived by calculating the free energy in a high temperature (weak coupling) 
expansion, with the first order calculation giving the mean field free energy, the second order the
TAP free energy and so on. For non-equilibrium and kinetic spin glass models, soft spin
systems were the first ones to be analyzed, using the Martin-Siggie-Rose generating functional 
formalism \cite{Sompolinsky81}. Spin glass models with hard spins were first treated in
\cite{SommersPRL87,Rieger88,Crisanti88}. A powerful generating functional approach was then 
developed by Coolen and collaborators \cite{CoolenPRL93,CoolenJPhysA94}, and it
forms the basis of our analysis here; see 
also \cite{FischerHertz,Coolen00} for reviews of 
the techniques used in both soft and hard spin models.  However, dynamical TAP equations to
describe the kinetics of order parameters for a specific 
realization of the disorder have been only derived for the spherical $p$-spin model \cite{Biroli99}
and the stationary state of the Ising spin model with asynchronous update 
dynamics \cite{Kappen00}.

In the same way that studying the quenched averaged kinetics of hard spin models usually
involves a different approach compared to soft spin models, deriving the dynamical TAP 
equations for hard spin models is somewhat different from doing so for their soft spin counterparts.
The aim of this paper is to develop a dynamical mean field theory
that relates the dynamics of mean magnetizations, potentially time varying
external fields, and the quenched couplings for two kinetic versions of
the Sherrington-Kirkpatrick model, one with synchronous update, the other
with asynchronous update. Using a generating functional approach, we 
derive the dynamical naive mean-field and TAP equations as first and second orders of
a high temperature expansion, similar to the equilibrium case for these two kinetic models.

In addition to the technical issues, the recent use of hard spin models with discrete states, 
e.g. Potts and Ising models \cite{Lezonetal06,WeigtPNAS2009,Cocco09,Roudi09-2}, in reconstructing 
the connectivity of biological networks encourages the study of the dynamics of these models 
in more detail. Once the forward dynamics is described, it is possible to use the results to 
construct approximations at the corresponding levels for the inverse problem: finding the 
couplings, given the magnetizations and correlation functions.  In this way, one can develop 
effective approximate reconstruction techniques that exploit the temporal structure of data and 
significantly improve the quality of network reconstruction in biological systems.
In fact, the results of this paper have been recently 
used in two other recent papers on the inverse problem \cite{Roudi11,Zeng10}.

The paper is organized as follows. After defining the dynamical models in the 
following section, we derive dynamical naive mean-field equations using the
generating functional for the synchronous updated model.  We report
the TAP equations, for which the details of the derivations are reported in the
Appendices. We then numerically calculate the errors for these kinetic equations as a function of
the strength of the couplings for the synchronous dynamics.

\section{Dynamical Model}
We consider a system of $N$ Ising spins $s_i=\pm 1, i=1,\cdots,N$
and assume that its state at time $t$,  $\bols(t)=\{s_1(t),\dots,s_N(t)\}$,
follows one of the following dynamics:
\begin{enumerate}
\item
{\bf Synchronous dynamics.} In this case time is discretized and 
the probability of being in state $\bols$ at time step
$t$, $p_t(\bols)$, is given by 
\numparts
\begin{eqnarray}
p_t (\bols)=\sum_{\bols'} W_t[\bols;\bols'] p_{t-1}(\bols')\label{DynSK-syna}\\
W_t[\bols;\bols']=\prod_i \frac{\exp(s_i \theta_i(t-1))}{2\cosh(\theta_i(t-1))}\label{DynSK-synb}\\
\theta_i(t)=h_i(t)+\sum_j J_{ij} s'_j(t)\label{theta-def} \label{DynSK-sync},
\end{eqnarray}
\endnumparts
This is, in other words, a Markov chain with transition probability $W_t$.
\item
{\bf Asynchronous dynamics.} In this case time is continuous and, $p_t(\bols)$
satisfies the following equation
\numparts 
\begin{eqnarray}
\frac{d}{dt}p_t(\bols)=\sum_i [p_t(F_i \bols) w_i(F_i \bols;t)-p_t(\bols) w_i(\bols;t)] \label{eq:synch_def_der} \label{DynSK-asyna}\\
w_i(\bols;t)=\frac{1}{2}[1-s_i \tanh[\theta_i(\bols;t)]] \label{DynSK-asynb},
\end{eqnarray}
\endnumparts
where the operator $F_i$ acting on $\bols$ flips its $i$th spin.
\end{enumerate}
For each of these processes one can define a generating functional. For the synchronous
case it takes the form of
\begin{equation}
Z[\psi,h]=\left \langle \exp \Big [\sum_{i,t} \psi_i(t) s_i(t)\Big ] \right \rangle,
\label{Z-def}
\end{equation}
\noindent where for any quantity ${\cal A}$ defined
as a function of a path $(\bols(T),\dots,\bols(0))$, $\langle \cdots \rangle$ 
indicates averaging over the paths 
taken by $\bols(t)$ under the stochastic dynamics of Eqs.\ \eref{DynSK-syna} --
\eref{DynSK-sync}, i.e.
\begin{equation}
\fl \langle {\cal A}\rangle =\Tr \ W_{T-1}[\bols(T);\bols(T-1)] \cdots W_0[\bols(1);\bols(0)]\ p_0(\bols(0))\ {\cal A}(\bols(T),\dots,\bols(0)),
\label{eq:angle_def} 
\end{equation}
and
\begin{equation}
\Tr \equiv \sum_{\bols(T)} \sum_{\bols(T-1)} \cdots \sum_{\bols(0)}.
\end{equation}
The asynchronous 
case is similar expect that the sum over $t$ in Eq.\ \eref{Z-def} should 
be replaced by an integration; see \ref{App2}.

It is useful to rewrite
the generating functional by considering $\theta_i(t)$ for each spin 
and each time step as a free parameter, integrate over it, and make
sure that the definition Eq.\ \eref{theta-def} is satisfied by 
inserting an appropriate delta function. This yields
 \begin{eqnarray}
\fl Z[\psi,h]&=\int D\btheta \left \langle \exp \Big [\sum_{i,t} \psi_i(t) s_i(t)  \Big ] \right \rangle \prod_{i,t} \delta \Big(\theta_i(t)-h_i(t)-\sum_{j}J_{ij}s_j(t)\Big) \nonumber\\
\fl &=\int D\btheta \hat{\btheta} \left \langle \exp \Big [i\sum_{i,t}\hatth_i(t)\{\theta_i(t)-h_i(t)-\sum_{j}J_{ij}s_j(t)\}+\sum_{i,t} \psi_i(t) s_i(t) \Big ] \right\rangle
\label{eq:Z_delta_def}
\end{eqnarray} 
where $D \btheta=\prod_{i,t} d \theta_i(t)$ and $D \btheta \hat{\btheta}=\prod_{i,t} \frac{1}{2 \pi} d \theta_i(t) d \hatth_i(t)$.
Using Eq.\ \eref{eq:angle_def} in Eq.\ \eref{eq:Z_delta_def}, we get
\numparts 
\begin{eqnarray}
Z_{\alpha}[\psi,h]&=\int D\btheta\hat{\btheta}\ \Tr \exp [L_{\alpha}]\label{eq:Z-L-def}\\
L_{\alpha}=&\sum_{i,t}\{i\hatth_i(t) [\theta_i(t)-\alpha \sum_j J_{ij}s_j(t)]+s_i(t+1) \theta_j(t)-\log \cosh \theta_i(t)\nonumber\\
&-ih_i(t)\hatth_i(t)+\psi_i(t) s_i(t)\},
\label{eq:L-def}
\end{eqnarray}
\endnumparts
where the parameter $\alpha$ is introduced to control the magnitude of the couplings, 
as will become clear later.

The generating functional has the property that its derivatives with respect to $\psi$ and $h$
give the averages of the correlators involving the spins and auxiliary fields. In particular,
defining 
\begin{equation}
\langle {\cal A} \rangle_{\alpha} =\frac{\int D\btheta\hat{\btheta}\ \Tr {\cal A}\ \exp[L_{\alpha}]}{\int D\btheta\hat{\btheta}\ \Tr \exp[L_{\alpha}]}
\end{equation}
and using Eqs. \eref{eq:Z-L-def} and \eref{eq:L-def}, we can define $m_i(t)$ 
and $\hatm_i(t)$ as
\numparts
\begin{eqnarray}
-i\hatm_i(t)\equiv\frac{\partial \log Z}{\partial h_i(t)}=-i \langle \hatth_i(t) \rangle_{\alpha}\label{fixed-mag-fielda}\\
m_i(t)\equiv \frac{\partial \log Z}{\partial \psi_i(t)}=\langle s_i(t)  \rangle_{\alpha}.
\label{fixed-mag-fieldb}
\end{eqnarray}
\endnumparts
From Eq. \eref{eq:Z_delta_def}, we can see that the $\psi \to 0$ limit of 
$\hatm_i(t)$ is the expected value of the auxiliary field $\hatth_i(t)$
under the measure inside the integral of Eq. \eref{eq:Z_delta_def}. It is easy to show
that this average, similar to the soft spin case, is zero. The same limit for $m_i(t)$  gives us the mean 
magnetizations. We therefore have
\begin{eqnarray}
&\langle \hatth_i(t) \rangle=\lim_{\psi\to 0} \hatm_i(t)=0\\
&\langle s_i(t) \rangle=\lim_{\psi \to 0} m_i(t).
\label{eq:phys_solu}
\end{eqnarray}
For a detailed discussion about these and other dynamical 
processes on Ising spin models see \cite{Coolen00}.

To derive the dynamical mean-field and TAP equations, one first calculates 
the Legendre transform of the logarithm of the generating functional of the process
defined by Eqs.\ \eref{DynSK-syna} -- \eref{DynSK-sync}. In this dynamical case, the logarithm of the
generating functional plays the role of the Helmholtz free energy in the equilibrium 
statistical mechanics while its Legendre transform corresponds
to the Gibbs free energy. One then expands this dynamical Gibbs free energy 
around the zero couplings limit, similarly to the equilibrium case \cite{Plefka82}
and the soft spin model \cite{Biroli99}.
In the following, we do this for Ising spins up to linear order in the couplings
for the synchronous update and use it to derive the dynamical mean-field equations.
The details of how to proceed to the TAP for the synchronous 
and asynchronous dynamics are provided in the Appendices. 

\section{Outline of the derivation of the dynamical equations}

The Legendre transform of the logarithm of the generating functional with respect to
the real fields, $h_i$, and the auxiliary fields, $\psi_i$ reads
\begin{equation}
\fl\Gamma[\hatm,m]\equiv\log Z[\psi[\hatm,m],h[\hatm,m]]-\sum_{i,t} \psi_i[\hatm,m](t) m_i(t)
+i\sum_{i,t} h_i[\hatm,m](t) \hatm_i(t),
\label{eq:Gamma-def}
\end{equation}
where $\psi$ and $h$ are now treated as functions of $\hatm$ and $m$ through the following
equalities
\numparts
\begin{eqnarray}
&\frac{\partial\Gamma}{\partial m_i(t)}=-\psi_i[\hatm, m](t) \label{fixed-mag-field-lega}\\
&\frac{\partial \Gamma}{\partial \hatm_i(t)}=ih_i[\hatm,m](t) \label{h-hm} \label{fixed-mag-field-legb}
\end{eqnarray}
\endnumparts

Eqs.\ \eref{fixed-mag-field-lega} and \eref{fixed-mag-field-legb} together
with the definition of $\Gamma_{\alpha}$ in
Eq.\ \eref{eq:Gamma-def} imply Eqs.\ \eref{fixed-mag-fielda} and \eref{fixed-mag-fieldb}.
Using Eq.\ \eref{Z-def} in Eq.\ \eref{eq:L-def}, $\Gamma_{\alpha}$ can also be written as
\numparts
\begin{eqnarray}
\fl \Gamma_{\alpha}[\hatm,m]&=\log \int D\btheta\hat{\btheta}\ \Tr e^{\Omega_{\alpha}} \label{Gamma_def}\\
\fl \Omega_{\alpha}&=\sum_{i,t}\Big \{ i\hatth_i(t) [\theta_i(t)-\alpha \sum_j J_{ij}s_j(t)]+s_i(t+1) \theta_i(t)-\log \cosh(\theta_i(t))\\
\fl &-ih_i(t)[\hatth_i(t)-\hatm_i(t)]+\psi_i(t)[s_i(t)-m_i(s)] \Big \} \nonumber
\end{eqnarray}
\endnumparts

The idea now is that for $\alpha=0$ the generating functional and its
Legendre transform can be easily calculated, as the spins will be independent of each other.
For the generating functional we have
\begin{equation}
Z_0[\psi,h]=\prod_i \prod^{T}_{t=1} \frac{2\cosh[h_i(t-1)+\psi_i(t)]}{2\cosh(h_i(t-1))},
\end{equation}
and for the Legendre transform of $\log Z_0$ we have
\begin{eqnarray}
\fl \Gamma_{0}[\hatm,m]&=\sum_{i,t} [\log(2 \cosh(h^{0}_i(t)+\psi^{0}_i(t+1))
 -\log 2\cosh(h^{0}_i(t))\\
\fl &-\psi^{0}_i(t) m_i(t) +ih^{0}_i(t) \hatm_i(t) ], \nonumber
\end{eqnarray}
where $h^{0}$ and $\psi^{0}$ are the real and auxiliary fields for which
Eqs.\ \eref{fixed-mag-fielda} and \eref{fixed-mag-fieldb} are satisfied for given $m$ and $\hatm$ 
at zero coupling ($\alpha = 0$), i.e.
\numparts
\begin{eqnarray}
&m_i(t)=\tanh[h^{0}_i(t-1)+\psi^{0}_i(t)]\label{m-psia}\\
&-i\hatm_i(t)=\tanh[h^{0}_i(t)+\psi^{0}_i(t+1)]-\tanh[h^0_i (t)]
\label{m-psib}.
\end{eqnarray}
\endnumparts

This can be used to express $h^{0}$ and $\psi^{0}$ in terms of $m$ and $\hatm$ as
\numparts
\begin{eqnarray}
&h^0_i(t)=\tanh^{-1} M_i(t)\label{eq:psi_ma}\\
&\psi^0_i(t)=\tanh^{-1} m_i(t)-\tanh^{-1} M_i(t-1) \label{eq:psi_mb},
\end{eqnarray}
\endnumparts
where $M_i(t)=m_i(t+1)+i\hatm_i(t)$.

To calculate the integral on the right hand side of Eq.\ \eref{Gamma_def} for $\alpha=1$, we 
can expand $\Gamma_{\alpha}$ around $\alpha=0$ and eventually set $\alpha=1$. 
Using the fact that
\begin{equation}
\frac{\int D\btheta\hat{\btheta}\ \Tr {\cal A}\ \exp[\Omega_{\alpha}]}{\int D\btheta\hat{\btheta}\ \Tr \exp[\Omega_{\alpha}]}
=\frac{\int D\btheta\hat{\btheta}\ \Tr {\cal A}\ \exp[L_{\alpha}]}{\int D\btheta\hat{\btheta}\ \Tr
\exp[L_{\alpha}]}
=\langle {\cal A} \rangle_{\alpha},
\end{equation}
for the first derivative of $\Gamma_{\alpha}$ with respect to $\alpha$ we have
\begin{equation}
\frac{\partial\Gamma_{\alpha}}{\partial \alpha}=
\left \langle \frac{\partial \Omega}{\partial \alpha}\right \rangle_{\alpha},
\end{equation}
yielding
\begin{eqnarray}
\frac{\partial\Gamma_{\alpha}}{\partial \alpha}=&-i\sum_{ij,t} J_{ij} \langle \hatth_i(t)s_j(t) \rangle_{\alpha}\label{Gamma_first_der1}\\
&-i\sum_{i,t}\frac{\partial h^{\alpha}_i(t)}{\partial \alpha} \langle [\hatth_i(t)-\hatm_i(t)] \rangle_{\alpha} 
+\sum_{i,t}\frac{\partial \psi^{\alpha}_i(t)}{\partial \alpha}\langle [s_i(t)-m_i(t)]\rangle_{\alpha}.\nonumber
\end{eqnarray}

The last two terms in Eq.\ \eref{Gamma_first_der1} are zero because of Eqs.\ \eref{fixed-mag-fielda} and \eref{fixed-mag-fieldb}; hence
\begin{equation}
\frac{\partial \Gamma_{\alpha}}{\partial \alpha}=-i\sum_{ijt} J_{ij} \langle \hatth_i (t) s_j(t)\rangle_{\alpha}.
\label{Gamma_first_der}
\end{equation}
The correlation function $\langle \hatth_i(t)s_j(t) \rangle_{\alpha}$ can also be
easily calculated at $\alpha=0$ yielding
\begin{eqnarray}
\fl-i\langle \hatth_i(t)s_j(t) \rangle_{0}=&\frac{1}{Z_0}\frac{\partial^2 Z_0}{\partial h_i(t)\partial \psi_j(t)}\nonumber\\
\fl &=(\tanh[h^0_i(t)+\psi^0_i(t+1)]-\tanh(h^0_i(t)]) \tanh[h^0_j(t-1)+\psi^0_j(t)],\nonumber\\
\fl &=-i \hatm_i(t) m_j(t)
\end{eqnarray}
where the last equality follows from Eqs.\ \eref{m-psia}--\eref{m-psib}.
Consequently, to first order in $\alpha$, we have
\begin{eqnarray}
\fl \Gamma_{\alpha}[\hatm,m]&=
\sum_{i,t} [\log 2\cosh(h^{0}_i(t-1)+\psi^{0}_i(t))
-\log 2 \cosh(h^{0}_i(t))]\nonumber\\
\fl &-\sum_{i,t} \psi^{0}_i(t)m_i(t)
+\sum_{i,t} i h^{0}_i(t) \hatm_i(t)
-i \alpha\sum_{i,j,t} J_{ij} \hatm_i(t)m_j(t)
\label{eq:Leg-sm-alpha}
\label{Gamma_first_order}
\end{eqnarray}
which combined with Eqs.\ \eref{eq:psi_ma} -- \eref{eq:psi_mb} gives
\begin{eqnarray}
\fl \Gamma_{\alpha}[\hatm,m]=&-\frac{1}{2}\sum_{i,t} \Big \{\log \Big[\frac{1+m_i(t)}{2}\Big]+\Big\{\log \Big [\frac{1-m_i(t)}{2}\Big ] \Big\}\\
\fl &+\frac{1}{2} \sum_{i,t} \Big\{ \log \Big [\frac{1+M_i(t)}{2}\Big]+\log \Big [\frac{1-M_i(t)}{2}\Big ]\Big\}\nonumber\\
\fl &-\sum_{i,t} m_i(t)\ \tanh^{-1}[m_i(t)]+ \sum_{i,t} M_i(t)\ \tanh^{-1}[M_i(t)] -i\alpha \sum_{i,j,t} J_{ij} \hatm_i(t) m_j(t)\nonumber\\
\fl &+\order(\alpha^2)\nonumber
\end{eqnarray}
Using Eq.\ \eref{fixed-mag-field-legb} yields
\begin{equation}
\tanh^{-1}M_i(t)=h_i(t-1)+\sum_{j} J_{ij} m_j(t-1).
\end{equation}
In the limit $\psi\to 0$ 
for which Eq.\ \eref{eq:phys_solu} is satisfied, we have
\begin{equation}
m_i(t+1)=\tanh \left[h_i(t)+\sum_{j} J_{ij} m_j(t)\right].
\label{dynSynchMF}
\end{equation}
This is the dynamical (naive) mean-field equation for the evolution
of the mean magnetization. The TAP equations can be derived in a
similar way by expanding $\Gamma_{\alpha}$ to second order in
$\alpha$, as shown in \ref{App1}. This yields the dynamical TAP 
equations
\begin{equation}
\fl m_i(t+1)=\tanh \left[h_i(t)+\sum_{j} J_{ij} m_j(t)-m_i(t+1) \sum_j J^2_{ij} [1-m_j(t)^2]\right].
\label{dynSynchTAP}
\end{equation}

To find the time evolving magnetizations for given external field and
coupling within the TAP approximation, the above equation should be solved
self consistently for $m_i(t+1)$ at each time step.  Note the form of the Onsager 
correction (the last term in Eq. \eref{dynSynchTAP}).  The $(1-m_j^2)$ term is 
evaluated at time step $t$, but $m_i$ is evaluated at time step $t+1$.  Thus \eref{dynSynchTAP}
is a set of equations to be solved for $m_i(t+1)$, not just a simple 
expression for $m_i(t+1)$ in terms of the $m_j(t)$, as in naive mean field theory.

The derivations of dynamical naive mean-field and TAP equations for the case of asynchronous dynamics
defined in Eqs.\ \eref{DynSK-asyna} and \eref{DynSK-asynb} are given in \ref{App2}. As shown there, 
these equations read

\begin{eqnarray}
\fl m_i(t)+\frac{d m_i(t)}{dt}=\tanh \left [ h_i(t)+\sum_j J_{ij} m_j(t)\right]\\
\fl m_i(t)+\frac{d m_i(t)}{dt}=\tanh \left [ h_i(t)+\sum_j J_{ij} m_j(t)-\Big (m_i(t)+\frac{d m_i(t)}{dt}\Big )
\sum_j J^2_{ij} (1-m^2_j(t))\right]\label{dynASynchTAP}
\end{eqnarray}

\section{Numerical results}

To test the dynamical naive mean field (hereafter: nMF) and TAP equations \eref{dynSynchMF} and
\eref{dynSynchTAP}, we ran simulations in which we simulated the process
define by \eref{DynSK-syna}-\eref{DynSK-sync} for $L$ time steps, for couplings drawn from 
a zero mean Gaussian distribution with variance $g^2/N$ ($J_{ij}$ is drawn independent of $J_{ji}$)
and subjected to two alternative types of external field. One
was a temporally constant field with a magnitude drawn independently for each spin
from a zero mean, unit variance Gaussian distribution. The other was a 
sinusoidally varying external field. For each sample of the $J$s and the fields,
we generated data from the system for $r$ repeats, calculated
$m_i(t)$ from these repeats, and used it in \eref{dynSynchMF} and
\eref{dynSynchTAP} to predict $m_i(t+1)$.  Finally, we calculated the
mean squared errors of these predicted values

\begin{equation}
{\rm MSE^{nMF/TAP}}=\frac{1}{L N}\sum^N_{i=1}\sum^L_{t=1} [m^{\rm nMF/TAP}_i(t+1)-m_i(t)]^2.
\label{MSE}
\end{equation}
The results for the two external fields used are shown below.

\subsection{Uniform field}

Fig.\ \ref{Fig1}A shows the dependence of the error for predicting
the magnetizations at time $t+1$, given the measured magnetizations at
$t$. Both TAP and nMF errors increase as $g$ increases, but the error
of nMF is always larger than that of TAP. Furthermore, how close to the true ($r \to \infty$)
values the measured magnetizations are systematically affects the nMF and TAP 
predictions: increasing $r$ decreases the errors for all $g$. This can also be seen in Fig.\ \ref{Fig1}B,
where the errors at $g=0.3$ are shown as functions of $r$, also
for two different values of $N$.

\begin{figure}[hbtp]
\center
\subfigure{\includegraphics[height=2.2in, width=2.2in]{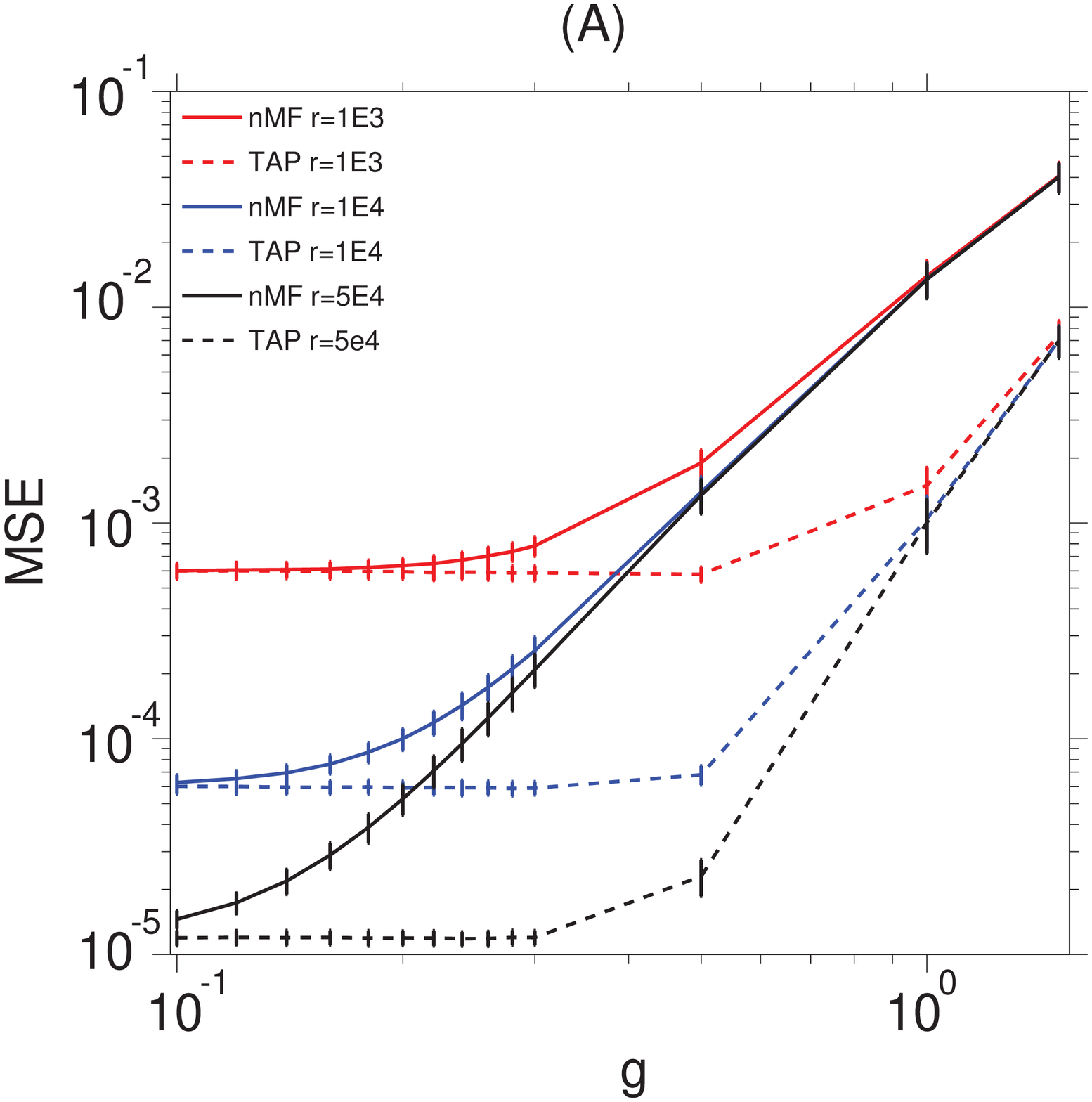}}
\subfigure{\includegraphics[height=2.2in, width=2.2in]{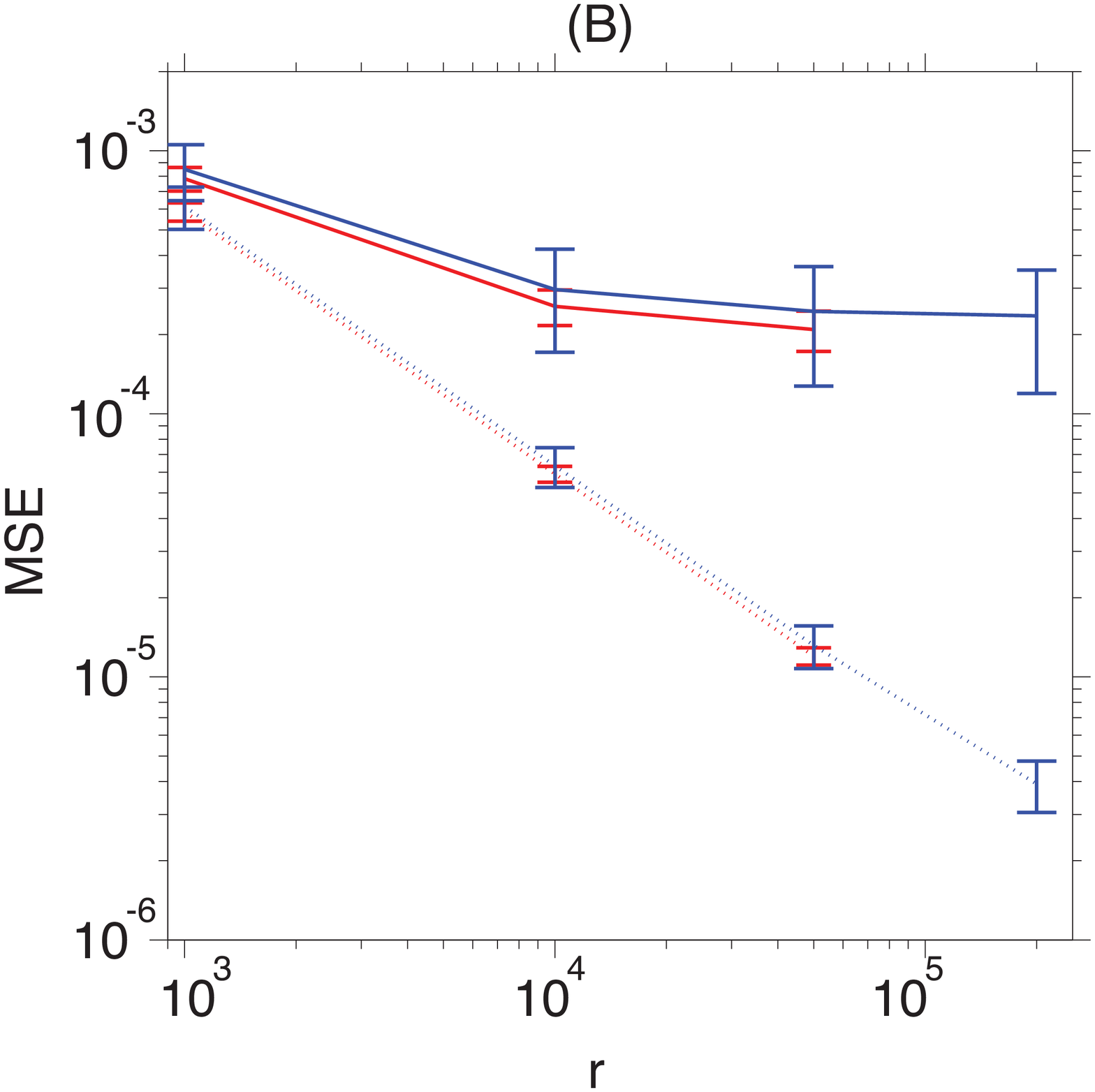}}\\
\subfigure{\includegraphics[height=2.2in, width=2.2in]{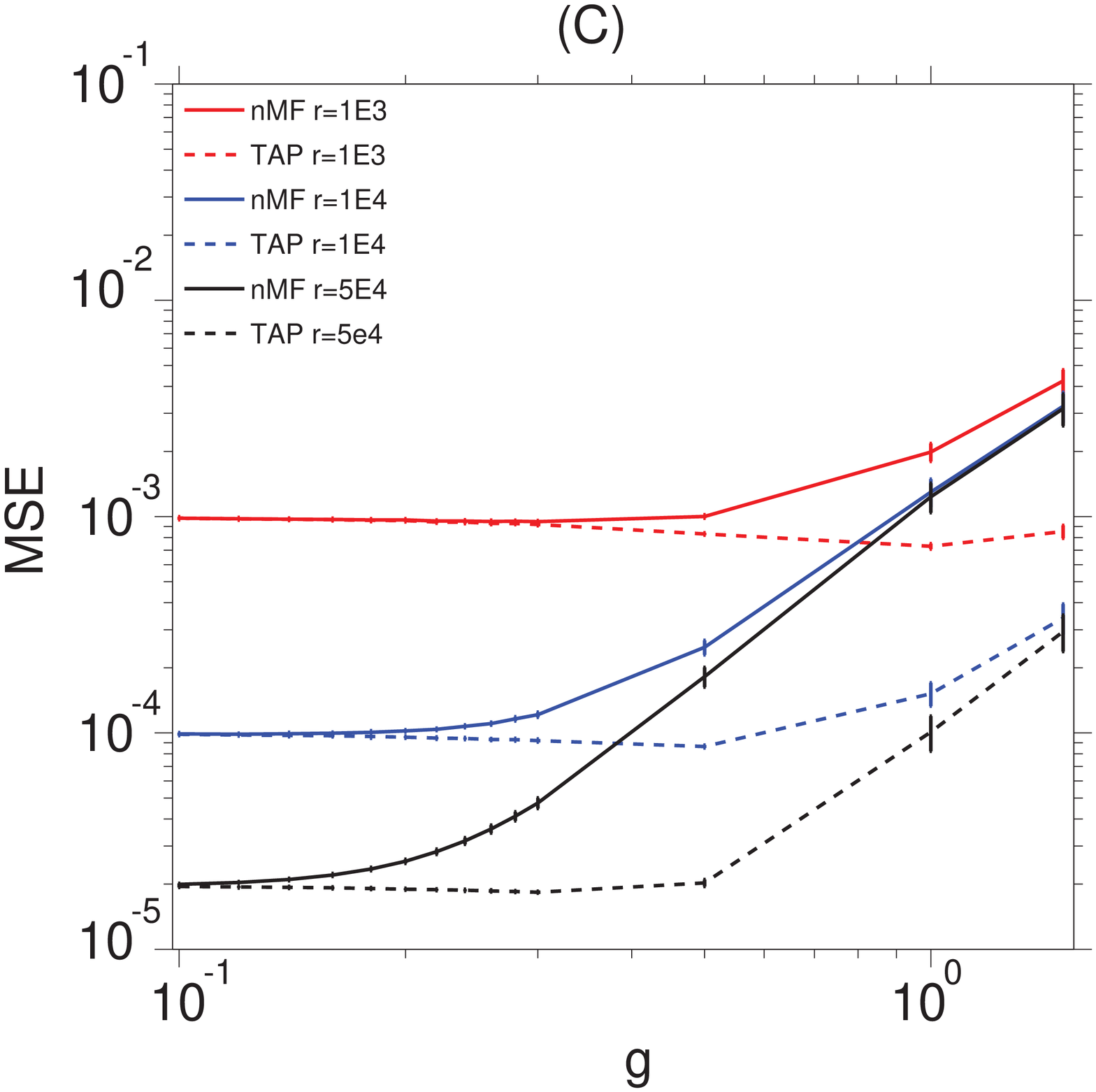}}
\subfigure{\includegraphics[height=2.2in, width=2.2in]{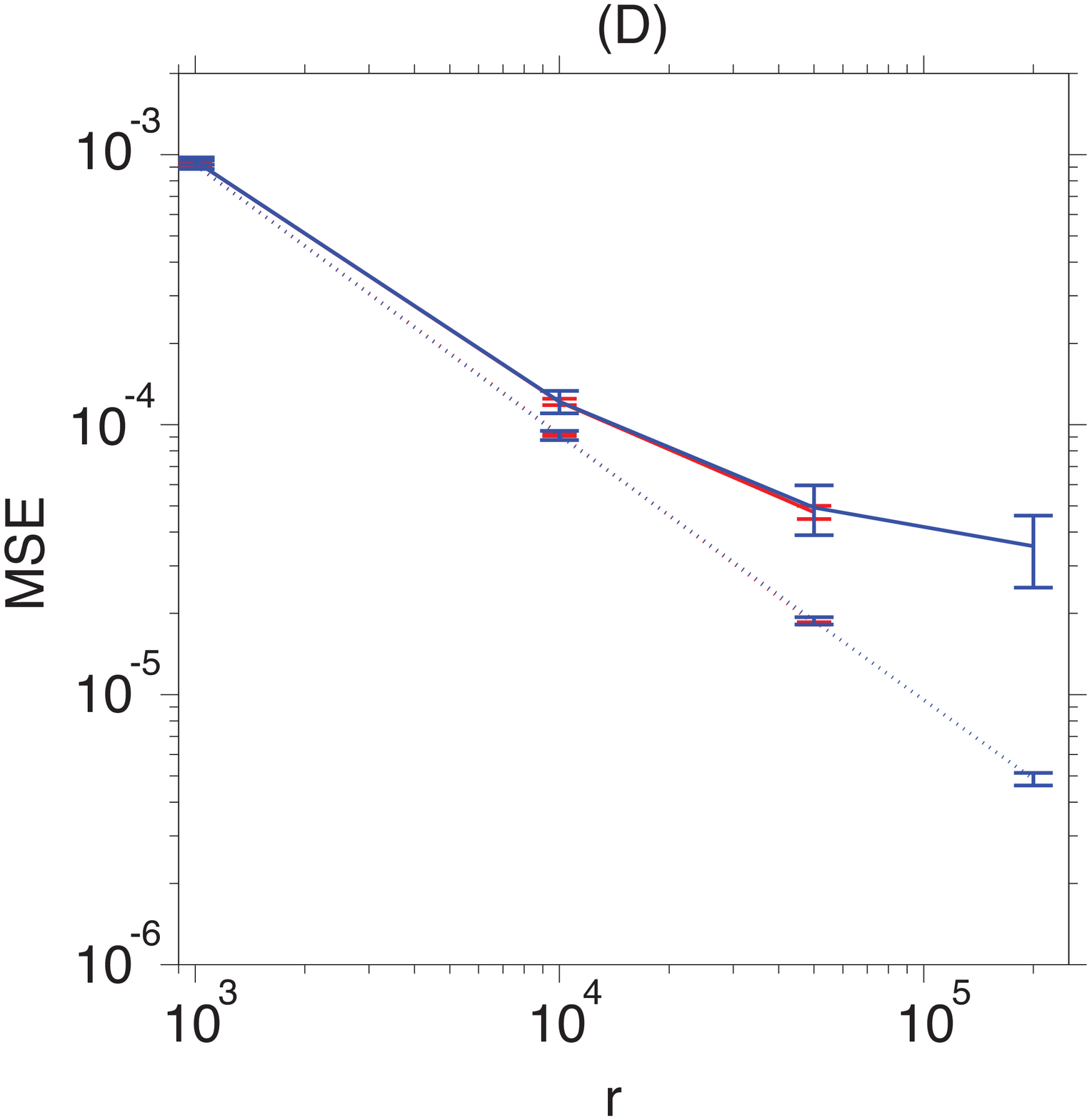}}\\
\caption{(A) The effect of magnitude of the couplings, $g$, on the
the error of TAP and nMF in predicting the magnetization at $t+1$ given the measured
magnetizations at $t$ using $r$ runs. (B) The effect of the number of runs, $r$, on the error
of TAP (red) and nMF (blue), for $N=10$ (full curve) and $N=50$ (dashed), and $g=0.3$. All errors are averages over
$25$ samples of the system. The errors bars show the standard deviation
of these samples. (C) and (D) the same as (A) and (B) but for a 
sinusoidal field.}
\label{Fig1}
\end{figure}

\subsection{Sinusoidal field}

Figs.\ \ref{Fig1} C and D show the same thing as Fig.\ \ref{Fig1} A and B, but
now the system is subjected to a sinusoidal external field 
with a peak amplitude of $0.1$ and a period of $20$ time steps. The results are 
qualitatively the same. For this case, we also look at the time dependence
of the errors in TAP and nMF equations. 

\begin{figure}[hbtp]
\center
\subfigure{\includegraphics[height=1.8in, width=1.8in]{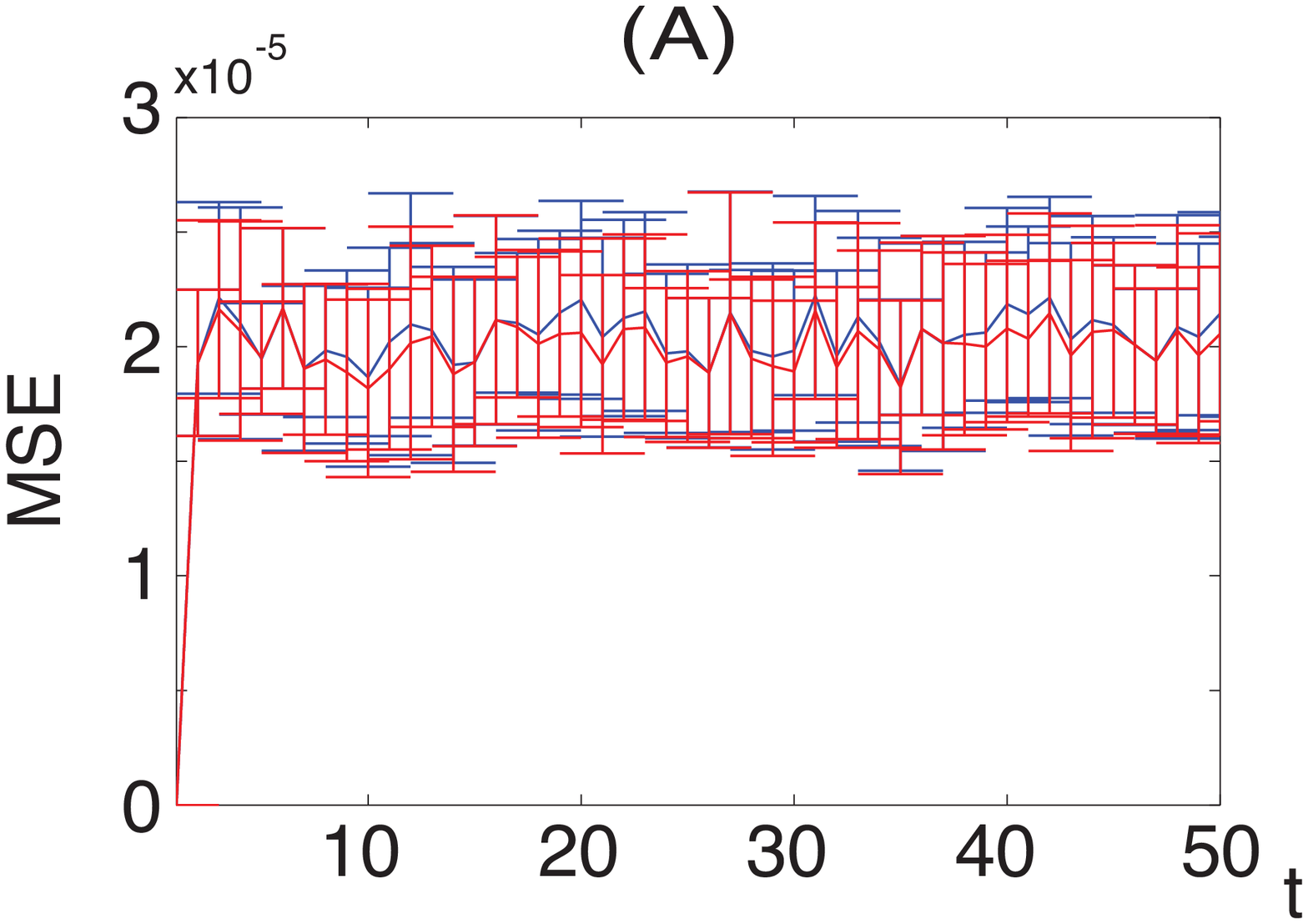}}
\subfigure{\includegraphics[height=1.8in, width=1.8in]{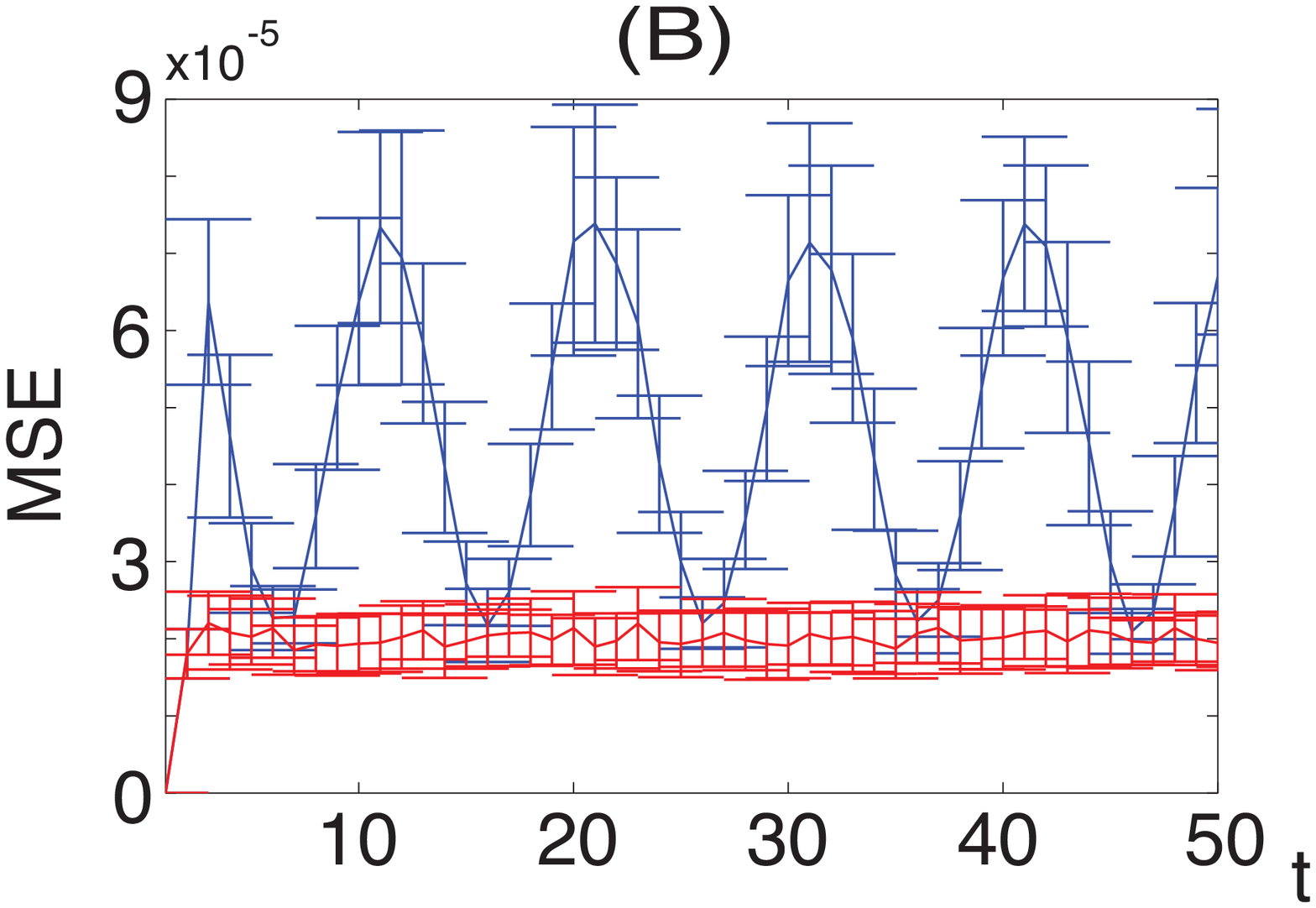}}
\subfigure{\includegraphics[height=1.8in, width=1.8in]{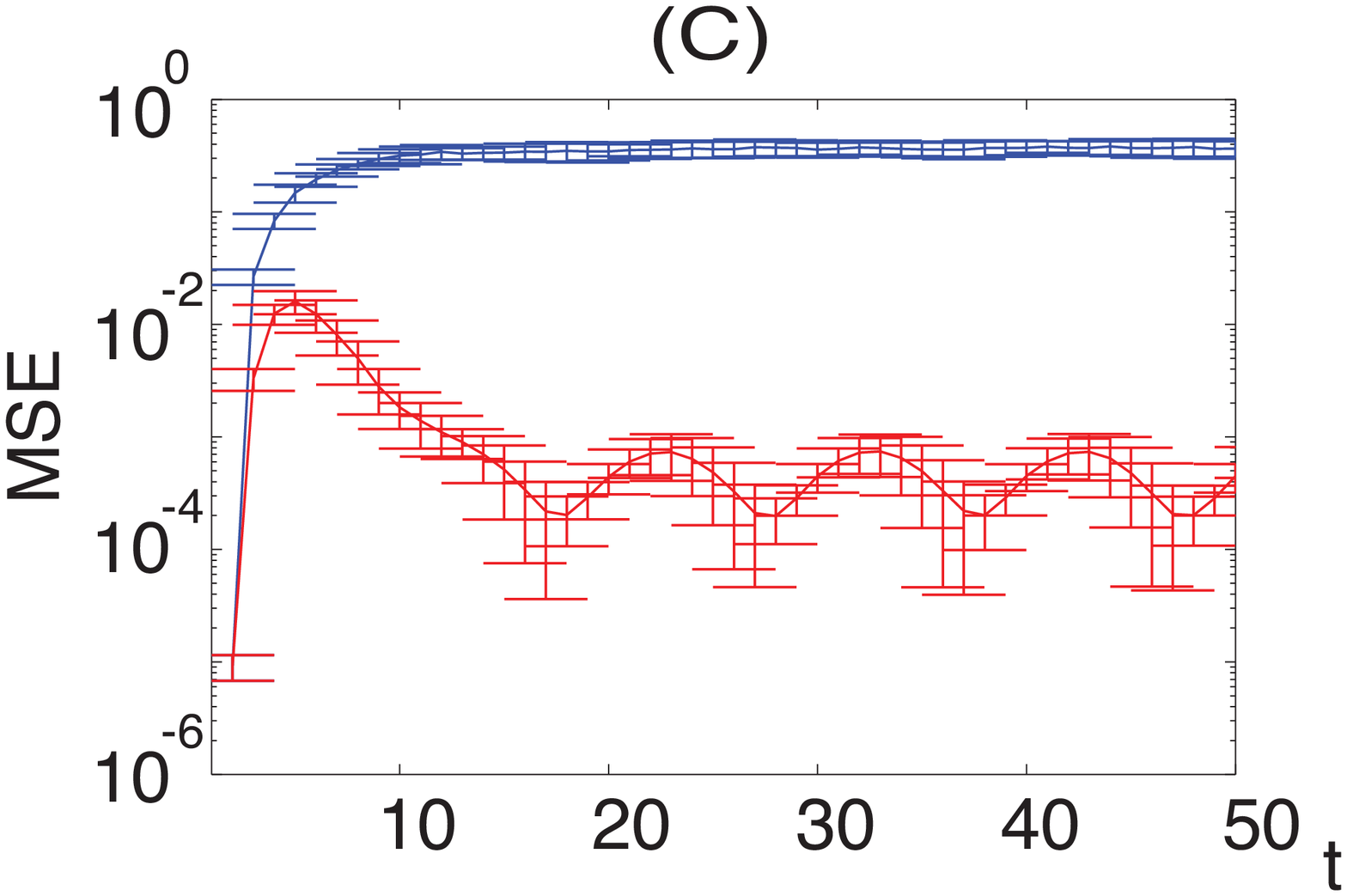}}\\
\subfigure{\includegraphics[height=1in, width=1.8in]{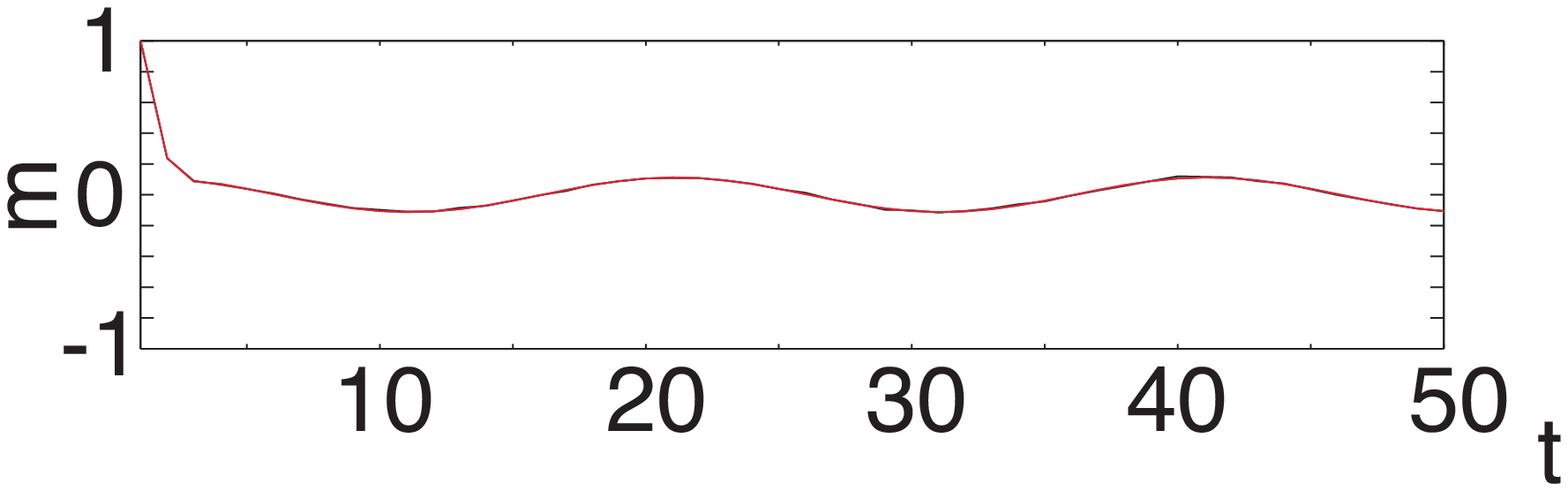}}
\subfigure{\includegraphics[height=1in, width=1.8in]{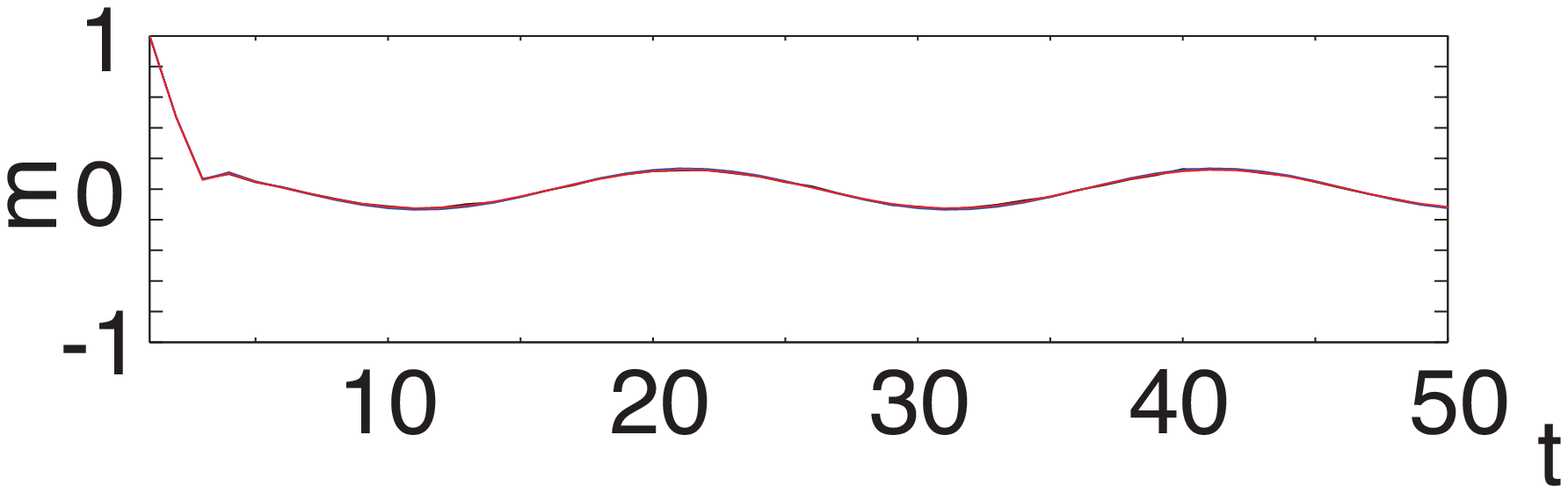}}
\subfigure{\includegraphics[height=1in, width=1.8in]{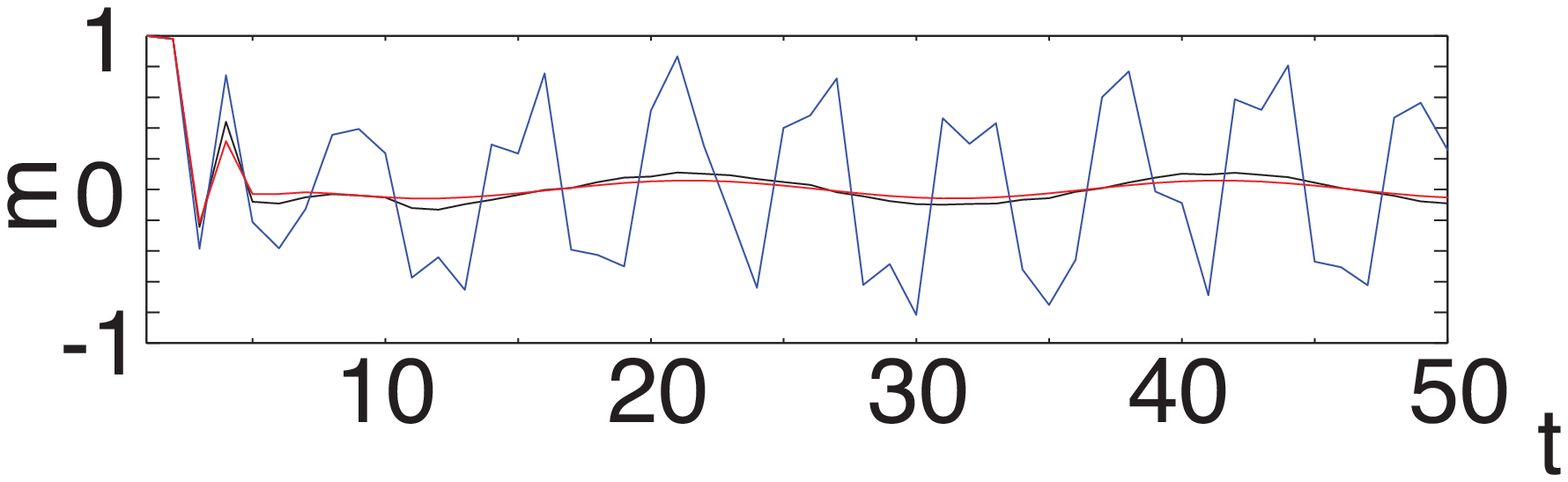}}
\caption{(A) Time dependence of the nMF (blue) and TAP errors (red) together with an
example of the measured magnetization (lower panel) from $50000$ repeats (black), nMF 
prediction (blue)
and TAP (red) for one spin and for $g=0.1$ and $N=40$. (B) and (C) 
show the same thing but for $g=0.28$ and $g=1.5$ respectively. Note the difference in the
scale of the y axes.}
\label{Fig2}
\end{figure}

Fig.\ \ref{Fig2} shows the time dependent
error (i.e. the right hand side of Eq.\ \eref{MSE} without averaging over time)
versus time. For weak coupling, the error of both nMF and TAP are very small. At intermediate
values of $g$, the error of nMF is still comparable to TAP. but fluctuating. At yet stronger
couplings, the nMF prediction very rapidly becomes different from the actual
measured values of the magnetizations.

\section{Discussion}
The TAP approach, formulated as a high temperature Taylor series
expansion of the equilibrium Gibbs free energy \cite{Plefka82}, is a powerful method
for studying equilibrium spin glass models. Similarly, 
dynamical TAP equations allow analyzing the dynamics of a single sample of a disordered 
system away from equilibrium. 
In this paper, we derived these equations for Ising spin glasses
with both synchronous and asynchronous updates. The main idea behind the 
derivation is similar to the one used by Biroli \cite{Biroli99} for 
the soft $p$-spin model obeying a Langevin equation,
with the difference that instead of a MSR formalism, we had to use
the generating functional approach of Coolen. For the $p$-spin model
the spherical condition results in the appearance of the the autocorrelation,
$\langle s_i(t) s_i(\tp) \rangle$, and response functions as order
parameters in dynamical TAP. For the hard spin 
Ising model, this is not the case. The response function can, of course,
be directly calculated from its definition and the TAP equations,
but calculating correlations, $\langle s_i(t) s_j(\tp) \rangle$,
function requires a different approach.

The derivation does not rely on the symmetry of the couplings 
and can, therefore, be applied to systems without detailed balance.
For the stationary case, the TAP equations are identical 
to those derived for the equilibrium model with symmetric connections.
This has been previously shown by Kappen and Spanjer \cite{Kappen00}
using an information geometric derivation
for the stationary state of the asynchronously updated model.
Numerical simulations with both a constant external field and a rapidly
evolving one show that the TAP equations predict the dynamics of the
individual site magnetizations very well. This may not be surprising
given the fact that the model we studied here was a kinetic variant of 
the SK model for which the equilibrium TAP equations provide the exact picture.

It is intriguing that the Onsager term in Eqs. \eref{dynSynchTAP} 
and \eref{dynASynchTAP} does not get the form $J_{ij} J_{ji} (1-m^2_j)$,
as would be expected from a simple reaction argument. This observation
has also been made earlier by Kappen and Spanjer \cite{Kappen00}.
A naive argument showing that the true correction to the mean-field equations is 
of the type $J^2_{ij} (1-m^2_j)$ is as follows. Starting from the 
exact equation
\begin{equation}
m_i(t+1)=\langle \tanh[h_i(t)+\sum_j J_{ij} s_j(t)] \rangle,
\end{equation}
we expand $\tanh$ around $b_i(t)=h_i(t)+\sum_j J_{ij} m_j(t)$ to
quadratic order in $\sum_j J_{ij} \delta s_j(t)$ where 
$\delta s_j(t)=s_i(t)-m_i(t)$. The linear term vanishes,
and using $\langle [\delta s_j(t)]^2\rangle=1-m^2_j(t)$ we
have
\begin{eqnarray}
m_i(t+1)=\tanh[b_i(t)] - (1-\tanh^2[b_i(t)]) \tanh[b_i(t)] \langle[\sum_{j} J_{ij} \delta s_j(t)]^2\rangle\nonumber\\
=\tanh[b_i(t)]-(1-\tanh^2[b_i(t)]) m_i(t+1) \sum_j J^2_{ij} (1-m^2_j(t))\nonumber\\
=\tanh[h_i(t)+\sum_j J_{ij} m_j(t)-m_i(t+1) \sum_j J^2_{ij} (1-m^2_j(t))]\nonumber
\end{eqnarray}
where in the second line we have used the mean field
equation $m_i(t+1)\approx \tanh(b_i)$.

An important issue that we have left out in this paper is the expected number
of solutions to the TAP equations for arbitrary couplings. It has been known for a long time
that, at low temperature, the expected number of solutions of the 
TAP equations for the SK model with symmetric couplings is 
exponential in $N$ \cite{Bray80}. It is also possible to calculate
the number of metastable states for couplings 
with an antisymmetric component at zero temperature \cite{Crisanti88}.
The TAP equations presented here allow extending the calculation in 
\cite{Bray80} to the type of couplings considered in \cite{Crisanti88} 
for non-zero temperatures. This calculation will be presented elsewhere.

The equilibrium TAP equations, derived for spin glass models with symmetric
couplings, can be used in deriving efficient approximations for solving the
inverse problem of reconstructing a spin glass model from samples of its
states \cite{Kappen98,Tanaka98}. As has been recently shown 
\cite{Roudi11,Zeng10}, the dynamical equations derived here can be
employed for taking the reconstruction to a more powerful level, allowing
for the reconstruction of systems outside equilibrium.

\ackn

The authors thank Erik Aurell
and Bert Kappen for discussions at various stages of this work. The use of computing resources
at Gatsby Computational Neuroscience Unit is gratefully acknowledged.

\appendix
\section{TAP equations for synchronous update}
\label{App1}
For deriving the TAP equations, we note that
\begin{eqnarray}
\frac{\partial^2  \Gamma_{\alpha}}{\partial{\alpha}^2}
=\left\langle\frac{\partial^2{\Omega}} {\partial\alpha^2}\right\rangle_{\alpha}
+\left\langle \left[\frac{\partial{\Omega}}{\partial \alpha}\right]^2\right\rangle_{\alpha}
-\left[\left \langle \frac{\partial{\Omega}}{\partial \alpha}\right\rangle_{\alpha} \right]^2
\label{Gamma-sec}
\end{eqnarray}
The first term on the right hand side of the above equation is equal to zero. To calculate 
the next two terms, we use the Maxwell equations
\numparts
\begin{eqnarray}
&i\frac{\partial h^{\alpha}_{i}(t)}{\partial \alpha}
=\frac{\partial}{\partial \hatm_i(t)} \frac{\partial \Gamma_{\alpha}}{\partial \alpha}
=-i\sum_{j} J_{ij} m_j(t)\\
&\frac{\partial \psi^{\alpha}_{i}(t)}{\partial \alpha}
=-\frac{\partial}{\partial m_i(t)} \frac{\partial \Gamma_{\alpha}}{\partial \alpha}
=i\sum_{j} \hatm_j(t) J_{ji} 
\end{eqnarray}
\endnumparts
to write
\begin{equation}
\fl \frac{\partial \Omega}{\partial \alpha}=-i\sum_{ijt} J_{ij} \hatth_{i}(t) s_{j}(t)+i\sum_{ijt} J_{ij} [\hatth_i(t)-\hatm_i(t)] m_{j}(t)
+i\sum_{ijt} J_{ji} [s_i(t)-m_{i}(t)] \hatm_j(t).\label{omega-alpha}
\end{equation}
We are therefore interested in calculating
\begin{equation}
\frac{\partial^2  \Gamma_{\alpha}}{\partial{\alpha}^2}=\left \langle \left[ \delta \left( \frac{\partial \Omega}{\partial \alpha} \right) \right ]^2 \right \rangle_{\alpha} =
\left \langle \left( \frac{\partial \Omega}{\partial \alpha} - \left \langle  \frac{\partial \Omega}{\partial \alpha} \right \rangle_{\alpha} \right)^2 \right \rangle_{\alpha}
\label{eq:Gammap2}
\end{equation}
where
\begin{eqnarray}
\delta \left( \frac{\partial \Omega}{\partial \alpha} \right) =&-i\sum_{ijt}\hat \theta_i(t) J_{ij}s_j(t)
+ i\sum_{ijt}\delta \hat \theta_i(t) J_{ij} m_j(t)\nonumber\\ 
 &+ i\sum_{ijt} \hatm_i(t) J_{ij} \delta s_j(t) 
+ i\sum_{ijt} \hatm_i(t) J_{ij} m_j(t).
\label{eq:delta_omeg}
\end{eqnarray}
Defining $\delta s_j(t) = s_j(t) - m_j(t)$ and $\delta \hat \theta_i(t) = \hat \theta_i(t) - \hatm_i(t)$, 
this can be rearranged into the following form:
\begin{equation}
\delta \left( \frac{\partial \Omega}{\partial \alpha} \right) 
= -i\sum_{ijt} \delta \hat \theta_i(t) J_{ij}\delta s_j(t).
\end{equation}
Now it is simple to evaluate Eq.\ \eref{eq:Gammap2}
\begin{equation}
\left \langle \left[ \delta \left( \frac{\partial \Omega}{\partial \alpha} \right) \right ]^2 \right \rangle_{\alpha}
= -\sum_{ijti'j't'} \langle  \delta \hat \theta_i(t) J_{ij}\delta s_j(t)   \delta \hat \theta_{i'}(t') J_{i'j'}\delta s_{j'}(t') \rangle_{\alpha}
\end{equation}
The factors have to be paired and for the pair averages we use
\numparts
\begin{eqnarray}
&\langle (-i \delta \hat \theta_i (t))^2 \rangle_{\alpha} = \frac{\partial \log Z_0}{\partial h_i(t)^2} = -\hatm_i^2(t) +2i\hatm_i(t) m_i(t+1)\\
&\langle -i \delta \hat \theta_i (t-1)\delta s_i(t)) \rangle_{\alpha} = \frac{\partial \log Z_0}{\partial h_i(t-1) \partial \psi_i(t)} = 1 - m_i^2(t)\\
&\langle (\delta s_i(t))^2 \rangle_{\alpha} = \frac{\partial \log Z_0}{\partial \psi_i(t)^2} = 1 - m_i^2(t).
\end{eqnarray}
\endnumparts
The terms containing products of two averages of the form $\langle \delta \hatth \delta s \rangle$ vanish,
because one pair factor has to have $t' = t-1$ and the other has to have $t = t' -1$, which cannot be satisfied 
simultaneously.  This leaves
\begin{eqnarray}
\left \langle \left[ \delta \left( \frac{\partial \Omega}{\partial \alpha} \right) \right ]^2 \right \rangle_0 
&= -\sum_{ijj't} \langle  (\delta \hat \theta_i(t))^2 \rangle_0 J_{ij} J_{ij'}\langle \delta s_j(t)\delta s_{j'}(t) \rangle_0\\ 
&= \sum_{ijt}[-\hatm_i^2(t)+2i\hatm_i(t) m_i(t+1)] J^2_{ij} [1-m_j^2(t)]\nonumber
\end{eqnarray}

Using this to calculate $\Gamma_{\alpha}$ to the quadratic order in $\alpha$, 
differentiating with respect to $\hatm_j(t)$, and setting $\hatm_j=0$
yields the dynamical TAP equations \eref{dynSynchTAP}.

\section{Asynchronous Dynamics}
\label{App2}
In the asynchronous case the generating functional takes the form
\begin{equation}
\fl Z_{\alpha}[\psi,h]=\int D\btheta\hat{\btheta} 
\prod_i \left \langle  \exp\Big [i\int dt \hat{\theta}_i(t) [\theta_i(t)-h_i(t)-\alpha \sum_{j} J_{ij}s_j(t)]
+\int dt \psi_i(t) s_i(t)\Big] \right \rangle
\end{equation}

\noindent and $\langle \cdots \rangle$ now indicates averaging with respect to the 
distribution defined by the solution to the differential equation Eq. \ \eref{eq:synch_def_der}.
This solution can be written as
\numparts
\begin{eqnarray}
p_t(\bols)=\prod_{i}\left [ \frac{1+\mu_i(t)}{2}\delta_{s_i(t),1}+\frac{1+\mu_i(t)}{2}\delta_{s_i(t),-1}\right]\\
\frac{d \mu_i}{dt}= - \mu_i + \tanh\theta_i(t), \ \ \mu_i(0)=s_i(0). \label{eq:u-def}
\end{eqnarray}
\endnumparts
The solution to Eq.\ \eref{eq:u-def} can be written as
\begin{equation}
\mu_i(t)=\int^{t}_0 d\tp\ e^{\tp-t} \tanh(\theta_i(\tp)) + e^{-t} \mu_i(0).
\label{eq:u-def2}
\end{equation}

The dynamical Gibbs free energy (i.e. the Legendre transform of the log generating functional)
is then
\begin{eqnarray}
\fl \Gamma_{\alpha}[\hatm,m]=&\log \int D\btheta\hat{\btheta} \prod_i 
\left \langle \exp\Big [i\int dt \hat{\theta}_i(t) [\theta_i(t)-h_i(t)-\alpha \sum_{j} J_{ij}s_j(t)]
+\int dt \psi_i(t) s_i(t)\Big ] \right \rangle \nonumber\\
\fl &+i \sum_i  \int dt\ h_i(t) \hatm_i(t)-\sum_i \int dt\ \psi_i(t) m_i(t)
\end{eqnarray}

\subsection{nMF for asynchronous update}
As we did for the synchronous case, we first calculate the non-interacting ($\alpha=0$) generating functional
\numparts
\begin{eqnarray}
&\log Z_0=\sum_{i} \int dt \log [\cosh(\psi_i(t))+\mu^0_i(t) \sinh(\psi_i(t))]\\
&\mu^0_i(t)=\int^{t}_0 d\tp\ e^{\tp-t}\tanh(h_i^0(\tp)) + e^{-t} \mu_i(0) \label{eq:uj0}
\end{eqnarray}
\endnumparts
and
\begin{equation}
\fl \Gamma_0[\hatm,m]=\sum_{i} \int dt \left\{\log [\cosh(\psi^0_i(t))+\mu^0_i(t) \sinh(\psi^0_i(t))]- \psi^0_i(t) m_i(t)+ih^0_i(t) \hatm_i(t) \right\}
\end{equation}
where now $\psi^{0}$ and $h^{0}$ are functions of $m$ and $\hatm$ from the following equations
\numparts
\begin{eqnarray}
&\frac{\delta \log Z_0}{\delta \psi^{0}_i(t)}=\frac{\sinh[\psi^{0}_i(t)]+\mu^0_i(t)\cosh[\psi_i(t)]}{\cosh[\psi_i(t)]+\mu^0_i(t)\sinh[\psi_i(t)]}=m_i(t)    \label{m-asynch}\\
&\frac{\delta \log Z_0}{\delta h_i(t)}=\int d\tp \chi^0_i(\tp,t)
\frac{\sinh[\psi^{0}_i(\tp)]}{\cosh[\psi^0_i(\tp)]+\mu^0_j(\tp)\sinh[\psi^0_i(\tp)]}=-i \hatm_i(t)\label{mhat-asynch}
\label{mmhat}
\end{eqnarray}
\endnumparts
and
\begin{equation}
\chi^0_i(\tp,t)=\frac{\delta \mu^{0}_i(\tp)}{\delta h_i(t)}=\Theta(\tp-t) e^{t-\tp} (1-\tanh^2[h^{0}_i(t)]).
\end{equation}

For nMF, we need to calculate
the linear term in $\alpha$. This is
\begin{equation}
\frac{\partial \Gamma_{\alpha}}{\partial \alpha}=-i \sum_{ij}J_{ij} \int dt \langle h_i(t) s_j(t) \rangle_{\alpha} = -i \sum_{ij} J_{ij} \int dt \hatm_i(t) m_j(t) 
\end{equation}
where the last equality follows from
\begin{equation}
\langle \theta_i(t) s_j(t) \rangle_{\alpha}= \frac{i}{Z_0}\frac{\delta^2 Z_0}{\delta h_i(t) \delta \psi_jt)}=\hatm_i(t) m_j(t), \ \ \ i\neq j.
\end{equation}
Consequently, up to the linear term in $\alpha$, we have
\begin{equation}
\Gamma_{\alpha}[\hatm,m]=\Gamma_0[\hatm,m]- i\alpha \sum_{ij}J_{ij} \int dt \hatm_i(t)m_j(t)
\end{equation}
Using the fact that $\partial \Gamma_0 /\partial \hatm_i(t)=ih^{0}_i(t)$, we find that 
\begin{equation}
ih_i(t)=ih_i^{0}(t)-i\sum_j J_{ij} m_j(t)
\end{equation}
Together with the fact that for $\psi^0=0$, we have $m_i(t)=\mu^{0}_i(t)$, the mean-field equation
is
\begin{equation}
\frac{d m_i(t)}{dt}+m_i(t)=\tanh[h_i(t)+\sum_{j} J_{ij} m_j(t)]
\end{equation}

\subsection{TAP equations for asynchronous update}

To derive the TAP equations, we need to calculate the second derivative
of $\Gamma$ with respect to $\alpha$. Similar to the synchronous update
case, we have
\begin{equation}
\frac{\partial^2 \Gamma_{\alpha}}{\partial \alpha^2}=
-\sum_{iji'j'}J_{ij} J_{i'j'}  \int dt dt' \langle \delta \hatth_i(t) \delta s_j(t)\delta \hatth_{i'}(t')  \delta s_{j'}(t')\rangle_{\alpha}
\end{equation}
and the non-zero contributions come from pairing the terms inside the averages. Non-zero contributions come from
$\langle \delta \hatth_i(t)^2\rangle_{\alpha}$. A correlation function of 
the form $\langle \delta \hatth_i(t) \delta s_{j'}(t')\rangle_{\alpha}$ is nonzero for $t'<t$ but since it always
appears multiplied by $\langle\delta s_{j}(t) \delta \hatth_{i'}(t')\rangle_{\alpha}$, which is zero for $t'<t$,
it does not contribute to the final results. We therefore have
\begin{eqnarray}
\frac{\partial^2 \Gamma_{\alpha}}{\partial \alpha^2}\Big |_{\alpha=0}&=\\
&-\sum_{ij} J^2_{ij} \int dt\ \langle [\delta \hatth_i(t)]^2 \rangle_{0} \langle [\delta s_j(t)]^2 \rangle_{0}\nonumber\\ 
&-\sum_{ij} J_{ij}J_{ji} \int dt  \langle \delta \hatth_i(t) \delta s_i(t) \rangle_{0} \langle \delta \hatth_j(t) \delta s_j(t) \rangle_{0}
\end{eqnarray}
To evaluate the above expression we first note that 
\numparts
\begin{eqnarray}
\fl \langle [\delta s_j(t)]^2\rangle_{0}=\frac{\delta^2 \log Z_0}{\delta \psi_j(t)^2}=1-m^2_j(t)\\
\fl \langle [\delta \hatth_i(t)]^2\rangle_{0}=-\frac{\delta^2 \log Z_0}{\delta h_i(t)^2}
=-\int d\tp \frac{\delta \chi_i^0(\tp,t)}{\delta h_i(t)}\gamma_i(\tp)+\int d\tp [\chi_i(\tp,t) \gamma_i(\tp)]^2 \label{eq:deltathsq}\\
\fl\langle \delta \hatth_j(t) \delta s_j(t)\rangle_0=i\frac{\delta^2 \log Z_0}{\delta h_j(t) \delta \psi_j(t)} =0.
\end{eqnarray}
\endnumparts
where the last equality follows from $\delta \mu^0_j(t)/\delta h_j(t)=0$ and 
\numparts
\begin{eqnarray}
\fl\gamma^0_i(t)\equiv \frac{\sinh[\psi^0_i(t)]}{\cosh[\psi^0_i(t)]+\mu^0_i(t)\sinh[\psi^0_i(t)]} \label{gammadef}\\
\fl\frac{\delta \chi_i^0(\tp,t)}{\delta h_i(t)}=
\frac{\delta^2 \mu_i^0(\tp)}{\delta h^2_i(t)}=-2 \tanh[h^0_i(t)] [1-\tanh^2[h^0_i(t)]] e^{\tp-t}\Theta(\tp-t)\nonumber\\
\fl=-2 \tanh[h^0_i(t)] \chi^0_i(\tp,t) \label{eq:eta_sec_der}
\end{eqnarray}
\endnumparts
Using Eq.\ \eref{eq:eta_sec_der} in Eq.\ \eref{eq:deltathsq} gives
\begin{equation}
\langle [\delta \hatth_i(t)]^2\rangle_{0}=-2i\tanh[h^0_i(t)] \hatm_i(t)+\int d\tp [\chi^0_i(\tp,t) \gamma^0_i(\tp)]^2
\label{eq:deltathsq2}
\end{equation}

The dynamical Gibbs free energy can then be written as
\begin{equation}
\fl \Gamma_{\alpha}[\hatm,m]=\Gamma_0[\hatm,m]-i\alpha \sum_{ij} J_{ij} \int  \hatm_i(t) m_j(t)
-\frac{1}{2}\alpha^2 \sum_{ij} J^2_{ij}\int dt \langle [\delta \hatth_i(t)]^2\rangle_{0} (1-m^2_j(t))
\end{equation}
where in the last sum $\langle [\delta \hatth_i(t)]^2\rangle_{0}$ should be considered
as a function of $m$ and $\hatm$.

\subsubsection{Stationary case}
For the stationary case we have $h^0_i(t)=h^{s0}_i$ and we have to 
take $t\to \infty$. This gives
\numparts
\begin{eqnarray}
&\mu^0_j=\tanh(h^{s0}_j) \label{eq:uh}\\
&m_i=\frac{\mu^0_i+\tanh[\psi_i]}{1+\mu^{0}_j\tanh[\psi_i]}    \label{eq:m-u-psi}\\
&-i\hatm_i=(1-[\mu^0_i]^2) \frac{\tanh[\psi_i]}{1+\mu^{0}_i\tanh[\psi_i]}   \label{eq:hatm-u-psi}\\
&\int d\tp [\chi^0_i(\tp,t) \gamma^0_i(\tp)]^2 =  (-i\hatm_i)^2
\end{eqnarray}
\endnumparts

Using Eqs.\ \eref{eq:uh}, \eref{eq:m-u-psi}, \eref{eq:hatm-u-psi} yields
\begin{equation}
m_i+i\hatm_i=\tanh(h^{s0}_i)
\end{equation}
Consequently, for the stationary case, Eq.\ \eref{eq:deltathsq2} can be written as
\begin{equation}
\langle [\delta \hatth_i(t)]^2\rangle_{0}=-2 i m_i \hatm_i+\order(\hatm^2).
\end{equation}
and therefore
\begin{equation}
\frac{\partial^2 \Gamma_{\alpha}}{\partial \alpha^2}\Big |_{\alpha=0}=2i \sum_{ij} J^2_{ij} \hatm_i m_i (1-m^2_j)+\order(\hatm^2)
\end{equation}
Using this the TAP equations in the stationary case would be

\begin{equation}
\tanh^{-1}m_i =h_i+ \sum_{j} J_{ij} m_j-m_i\sum_{j} J^2_{ij} (1-m_j^2)
\end{equation}

\noindent which is identical to the result of \cite{Kappen00}

\subsubsection{General case}

Under general conditions we cannot express $h$ and $\psi$ explicitly in 
terms of $m$ and $\hatm$.  However, we can still calculate $\partial \Gamma/\partial \hatm_i(t)_i$
at $\hatm=0$, which is what we need for deriving the TAP equations.  

First note that the second term on the right-hand side of Eq.\ \eref{eq:deltathsq2} is of 
quadratic order in $\psi$ in the limit $\psi \to 0$ (from Eq.\ \eref{gammadef}).  But $\hatm$ is linear
in $\psi$ (from Eq.\ \eref{mhat-asynch}), so this term is of second order in $\hatm$ and 
its derivative with respect to $\hatm$ vanishes as $\hatm \to 0$.  Thus we can discard 
it in finding the TAP equations.

We are now interested in the following quantity
\numparts
\begin{eqnarray}
\fl \frac{\delta }{\delta \hatm_i(t)} \int d \tp \langle [\delta \hatth_j(\tp)]^2\rangle_{0} (1-m^2_k(\tp))=\nonumber\\
\fl
-2i x_j(t) (1-m^2_k(t)) \delta_{ij}-2i \int d\tp \frac{\delta x_i(\tp)}{\delta \hatm_i(t)} \hatm_{j}(\tp) (1-m^2_k(\tp)) 
\label{eq:theta-sq}
\end{eqnarray}
\endnumparts
where $x_i(t)=\tanh(h^0_i(t))$. For $\hatm_j(t) \to 0$,  The only term that will be nonzero 
on the right hand side of Eq.\ \eref{eq:theta-sq} is the first, as long $\delta x_i(\tp)/\delta \hatm_i(t)$ does not
diverge as fast as or faster than $1/\hatm$ as $\hatm \to 0$. Whether
$\delta x_i(\tp)/\delta \hatm_i(t)$ is regular in the limit $\hatm \to 0$ or not depends
on whether the functional matrix $\delta (m,\hatm )/\delta (h,\psi )$ is regular in this limit. 
The latter is not singular when the generating functional is regular unless the system is at a phase 
transition. Assuming that this is not the case, we can ignore the last
term in Eq.\ \eref{eq:theta-sq}.

Now we can proceed the way we did in the naive mean-field case, but evaluating $\Gamma_{\alpha}$ to 
second order in $\alpha$.  The functional derivative of
$\Gamma_{\alpha}$ with respect to $\hatm$, evaluated at $\alpha = 1$, gives $ih$:
\begin{equation}
ih_i(t) = ih_i^0(t) -i\sum_jJ_{ij}m_j(t) + i\tanh [h_i^0(t)]\sum_jJ_{ij}^2[1-m_j^2(t)],
\end{equation}
$\tanh h_i^0(t)$ can be related to $\mu_i^0(t)$ through
\begin{equation}
 \frac{d \mu_i^0}{dt}= - \mu_i^0 + \tanh h_i^0(t),
\end{equation}
and $\mu_i^0 \to m_i$ when $\psi$ and $\hatm \to 0$, yielding the TAP equations
\begin{equation}
\frac{d m_i(t)}{dt}+m_i(t)= \nonumber 
\end{equation}
\begin{equation}
\tanh\left[ h_i(t)+\sum_{j} J_{ij} m_j(t)-\left( \frac{dm_i(t)}{dt} + m_i(t)\right)\sum_jJ_{ij}^2[1-m_i^2(t)] \right]
\end{equation}
Note that these are of the same form as those for the synchronous-update model with $m_i(t+1)$ replaced by $m_i + dm_i/dt$.

\section*{References}
\bibliographystyle{unsrt}
\bibliography{mybibliography2010}
\end{document}